\documentclass[conference]{IEEEtran}
\IEEEoverridecommandlockouts

\usepackage{amsmath,amssymb,amsfonts}
\usepackage{algorithmic}
\usepackage{graphicx}
\usepackage{textcomp}
\usepackage{comment}
\usepackage{booktabs}
\graphicspath{{./images/}} 
\DeclareGraphicsExtensions{.pdf,.jpg,.png,.jpeg}

\usepackage[style=ieee,doi=false,isbn=false,url=false,eprint=false,maxnames=4]{biblatex}
\title{{\normalsize Author preprint July 2026}\\
Typical models of the distribution system restoration process
\thanks{Support from USA NSF grants 2153163 and 2429602, Argonne National Laboratory, and PSerc project S110 is  gratefully acknowledged.}}

\author{\IEEEauthorblockN{Arslan Ahmad}
\IEEEauthorblockA{Iowa State University\\
arslan@iastate.edu}
\and
\IEEEauthorblockN{Ian Dobson}
\IEEEauthorblockA{Iowa State University\\
dobson@iastate.edu}
}

\begin{document}
\maketitle	

\begin{abstract}
\looseness=-1
Accurate probabilistic modeling of the power system restoration process is essential for resilience planning, operational decision-making, and realistic simulation of resilience events. In this work, we develop data-driven probabilistic models of the restoration process using outage data from four distribution utilities. We decompose restoration into three components: normalized restore time progression, total restoration duration, and the time to first restore. The Beta distribution provides the best fit for restore time progression, and the Uniform distribution is a defensible, parsimonious approximation for many events. Total duration is modeled as a heteroskedastic Lognormal process that scales superlinearly with event size. The time to first restore is well described by a Gamma model for moderate and large events. Together, these models provide an end-to-end stochastic model for Monte Carlo simulation, probabilistic duration forecasting, and resilience planning that moves beyond summary statistics, enabling uncertainty-aware decision support grounded in utility data.
\end{abstract}


\section{Introduction}
Power restoration after an outage event is a complex time-evolving process in which customers and components are progressively restored as switching, repairs, and crew actions unfold. 
Yet more often, restoration is represented by a single summary statistic (e.g., an average restoration time) or by convenient assumptions (e.g., exponential repair), which obscures how restoration unfolds during an event and limits the modeling of restoration uncertainty in planning studies. 
Restoration process models matter because they 
(i) provide a probabilistic description of how restoration progresses in time rather than a single estimate, 
(ii) enable Monte Carlo simulations that require realistic restoration dynamics, and 
(iii) support restoration and resilience planning using data-driven models rather than ad hoc assumptions.

In this paper, we use unscheduled outage data from four distribution utilities to model the restoration process of resilience events as three components: (1) the normalized progression of restore times (restore process shape), (2) the total restoration duration, and (3) the time to first restore. We identify model classes that fit consistently across utilities under a common event definition, while distinguishing common restoration behavior from utility-specific effects.

\noindent
The questions we answer in this work are:
\begin{itemize}
    \item What probabilistic model class best represents restoration times of distribution system outage events across multiple utilities, under a consistent event definition, and with statistical evidence?
    \item What is stable vs utility-specific about restoration-time behavior?
    \item What do the ``best-fit" probabilistic models enable that common practice (means/medians, or ad hoc assumptions like exponential) do not?
\end{itemize}

\section{Related Work}
Most existing work on distribution-system restoration focuses on predicting outage duration or restoration time, often for individual outages or for specific hazard classes. For hurricane-related outages, statistical and machine-learning models have been compared for outage-duration prediction, with BART and log-linear regression reported as strong predictors 
\cite{NateghiRISK11, WillemsIEEEACCESS24}. More recent work has used transfer learning, deep neural networks, and natural-language processing of field reports to improve restoration-time prediction and real-time outage-duration estimates \cite{WangPS24, ArifPMAPS18, JaechPS19}. These approaches are valuable for forecasting restoration time, but they generally do not model the full temporal process of restoration within an event.

A second body of work models restoration and resilience processes more directly. Event-based resilience metrics have been developed for transmission-system events \cite{DobsonPS24}, and outage and restore process statistics have been extracted from distribution utility data to compute distribution-system resilience metrics \cite{CarringtonPS21}. Other studies have modeled distribution restoration using queuing and sequential Monte Carlo simulation \cite{ZapataTDEXPO08}, empirical restoration-time statistics and covariates \cite{ChowTPD96}, survival or accelerated-failure-time models \cite{JamalIJDRS23}, post-disaster repair scheduling \cite{TanPS19}, and data-driven restoration-duration models conditioned on weather zones \cite{WangNature26}.

In contrast to the existing literature, this work 1) focuses on modeling the complete restoration process, not just the restoration duration, 2) uses events instead of individual outages, 3) models a large number of all types of large events instead of just a few extreme events of a particular type, and 4) builds the models using outage data from multiple distribution utilities.
In summary, this work focuses on \textit{utility-data-driven} modeling of the \textit{restoration process} for \textit{events} in \textit{distribution systems}.

\section{Utility Data}
We use real-world data of unscheduled power outages from four different USA distribution utilities:
Utility 1 (anonymous),
Utility 2 (north central Massachusetts),
Utility 3 (western Massachusetts counties), and
Utility 4 (mainly Suffolk County, Massachusetts). The data from utilities 2, 3, and 4 is publicly available at \url{https://www.mass.gov/info-details/power-outages}.
Summary of the data is given in Table~\ref{table:dataSummary}.
We group individual outages into events, where an event is a set of outages that overlap in time.
We calculate the restore process of each event as the cumulative number of outages restored during the event.
More details about the events and processes are in \cite{CarringtonPS21,ahmadArxiv25}.

\begin{table}[hbpt]
	\caption{Utility Data} 
	\label{table:dataSummary}
	\centering
\begin{tabular}{ l @{\hspace{-1pt}} c c c c}
                        &Utility-1  &Utility-2  &Utility-3  &Utility-4  \\
\midrule
Time period (years)     & 6         & 11        & 10        & 11        \\
Total outages           & 32278     & 6371      & 22371     & 13340     \\
Total events            & 5716      & 3832      & 7000      & 6485      \\
\# of events of size $\geq$30 & 132       & 13        & 71        & 17        \\ 
\end{tabular}
\end{table}

\section{Methods}
Consider an event with $n$ outages. Let $r_1,r_2,...,r_n$ be the absolute restore times of outages sorted in ascending order, and $\Delta r_1,\Delta r_2,...,\Delta r_n$ be the normalized restore times of outages relative to the first restore time, i.e., $\Delta r_i = (r_i-r_1)/D$, with $\Delta r_i \in [0,1]$, and $D=r_n-r_1$ being the total duration of the restoration process.
Let the time to first restore in the event be $D_{r1}=r_1-o_1$, where $o_1$ is the start time of the first outage in the event.
Our goal is to model the typical restoration process \eqref{eq:RestorationProcess} by modeling $D$, $D_{r1}$, and $\Delta r_i$.
The absolute restoration times can then be recovered using
\begin{equation}
\label{eq:RestorationProcess}
    r_i = o_1+ D_{r1}+D\Delta r_i  \quad \text{for} \quad i=1,2,3,...,n
\end{equation}

\subsection{Modeling the Restore Times}
Once the restoration process starts, the normalized restore times $\Delta r_i$ in an event can be treated as an observed sample from a random variable $R$. We want to estimate the distribution of $R$ using the empirical data from many events. 
We evaluate the following candidate distributions to find a distribution that best fits the observations of $R$:
\begin{itemize}
    \item Lognormal, ${\rm Lognormal}(\mu,\sigma  ^2)$
    \item Exponential, ${\rm Exp}(\lambda),~\lambda >0$
    \item Uniform, $U(0,1)$
    \item Beta (with 1 parameter), ${\rm Beta}(\alpha,\alpha),~ \alpha >0$
    \item Beta (with 2 parameters), ${\rm Beta}(\alpha,\beta),~ \alpha,\beta >0$
\end{itemize}

\noindent We model the event restore times in two different ways:
\begin{enumerate}
    \item \textit{Global Model}: We normalize and combine the restore times of all the events together to make a single set of observations and fit different candidate distributions to that set. The results are shown in Figs.~\ref{fig:globalFitComparison} and \ref{fig:allUtilities}.
    Estimated parameters for the best-fit of each distribution using this pooled-fit method are shown in Table~\ref{table:globalResults}.
    \item \textit{Individual Model}: We fit candidate distributions to the restore times of each individual event and evaluate the number of cases in which one candidate distribution gives a better fit than the other. The results for these comparisons are given in Table~\ref{table:gofRestoreTimes}. This method gives us a range of fitted parameters for each distribution.
\end{enumerate}
The model parameters are estimated via Maximum Likelihood Estimation (MLE). Goodness-of-fit is evaluated qualitatively using the Cumulative Distribution Function (CDF) and quantile plots, and quantitatively using Kolmogorov-Smirnov (KS) distance.
For comparison between distributions, we use the AICc \eqref{eq:AIC} and the Likelihood Ratio Test (for nested models only) to determine the best-fitting model while accounting for model complexity. 
\begin{align}
    \label{eq:AIC}
    {\rm AICc} = {\rm AIC} + \frac{2k^2 + 2k}{n-k-1},\qquad {\rm AIC} = 2k-2{\rm ln}[\hat{L}]
\end{align}
Here $k$ is the number of parameters in the model, $n$ is the sample size, and $\hat{L}$ is the likelihood estimate. AICc adjusts the Akaike Information Criterion (AIC) for small sample sizes.

\subsection{Modeling the Event Restoration Duration}

\begin{figure*}[ht]
    \centering
    \includegraphics[width=\textwidth]{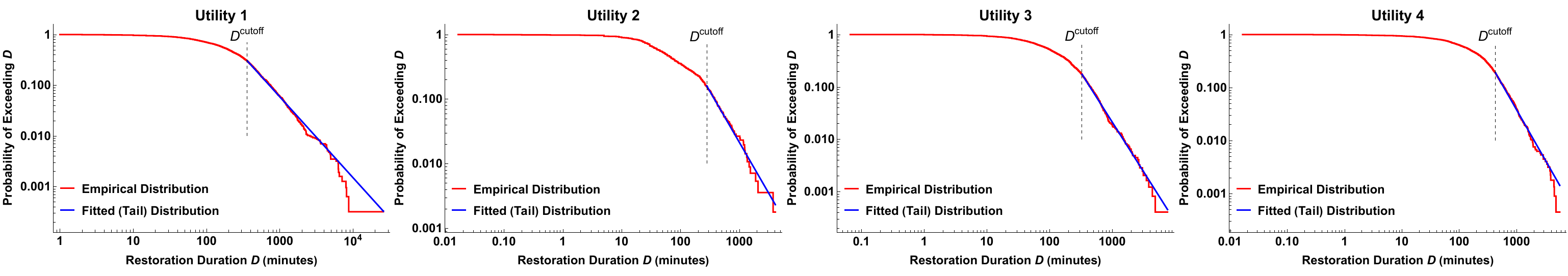}
    \caption{Empirical CCDF of Event Restoration Duration $D$ of events with $n \geq 2$. The straight line shows a Pareto fit to the distribution tail starting at $D^{\rm cutoff}$.}
    \label{fig:restorationDuration}
\vspace{-1em}
\end{figure*}

To model the restoration duration $D$ (in minutes) of events, we consider strictly positive values; hence, only events with 2 or more outages are considered, as single-outage events have a restoration duration of 0. We start by analyzing the empirical distribution of restoration duration across all utilities and observe heavy tails (see Fig.~\ref{fig:restorationDuration}). The slope magnitudes\footnote{The slope magnitudes $\alpha_{\rm p}$ and the cutoff values $D^{\rm cutoff}$ are determined automatically using Clauset's methods from \cite{ClausetSIAM09}} of the distribution's Complementary Cumulative Distribution Function (CCDF; equals 1-CDF) tails on log-log plots range from 1.6 to 1.9, as indicated in Table~\ref{table:restorationDuration}. 
For descriptive statistics, the distribution of restoration duration can be modeled using a spliced distribution where the tail is modeled using a Pareto distribution with parameter $\alpha_{\rm p}$ and cutoff values given in Table~\ref{table:restorationDuration}, and the body of the distribution is modeled using a Lognormal distribution.
However, a detailed model with a generative explanation is introduced below.

\begin{figure}[t]
    \centering
    \includegraphics[width=1.0\linewidth]{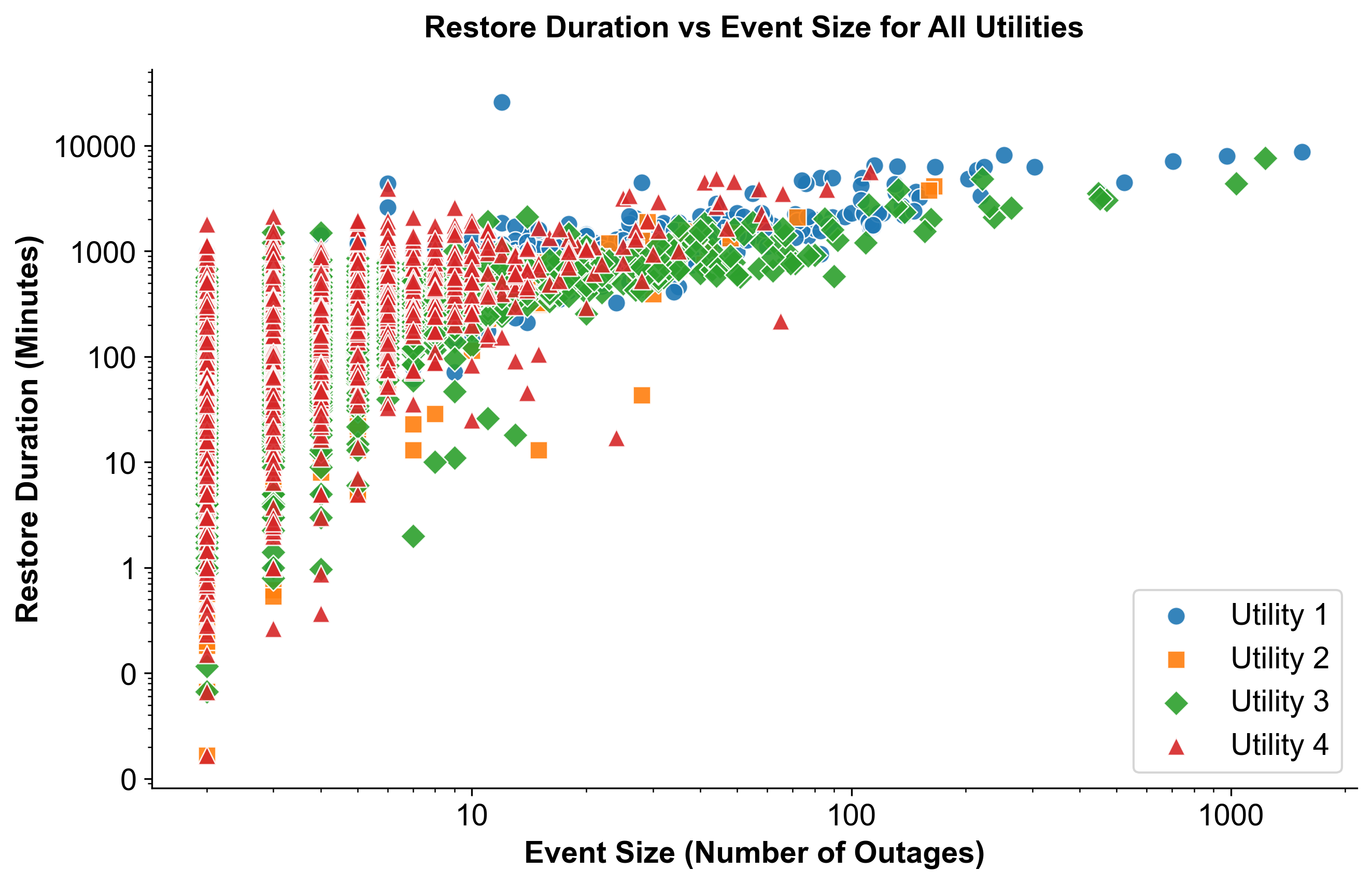}
    \caption{Event Restoration Duration versus Event Size (log-log scale).}
    \label{fig:restoreDurationVsEventSize}
\end{figure}

\begin{table}[hbpt]
	\caption{Fitted parameters of the right tail of the distribution of Event Restoration Duration} 
	\label{table:restorationDuration}
	\centering
\begin{tabular}{ l @{\hspace{-1pt}} c c c c c}
                        &Utility-1  &Utility-2  &Utility-3  &Utility-4  \\
\midrule
$D^{\rm cutoff}$      & 360       & 280       & 330       & 429         \\    
Slope Magnitude $\alpha_{\rm p}$& 1.60      & 1.58      & 1.89      & 1.90          \\   
\end{tabular}
\end{table}

Fig.~\ref{fig:restoreDurationVsEventSize} shows a scatter plot of event restoration duration $D$ and event size $n$ on a log-log scale, clearly showing that, for each fixed event size, restoration durations span orders of magnitude across all utilities\footnote{We find that event size and restoration duration have a statistically significant (p-value $<$ 0.05) positive correlation, with both Pearson and Spearman correlation coefficients exceeding 0.5.}.
This approximately linear relationship between restoration duration and event size on a log-log scale suggests power-law scaling, i.e., $D \propto n^\beta$.
This means that the restoration duration is not generated by a single process; rather, it is a conditional outcome dependent on the underlying event size, which has two regimes: small and large events.
These regimes differ in causes (equipment failure vs. systematic vulnerability), response protocols (local repair vs. coordinated restoration and prioritization), and resource requirements (single crew vs multiple crews and mutual aid). 
Therefore, we model the event restoration duration conditionally on the event size, explained as follows:
\begin{equation}
\label{eq:restorationDurationModel}
    D_n = D|N = n~~\sim~~{\rm Lognormal}(\mu(n),\sigma^2(n))
\end{equation}
We choose the lognormal distribution because of its positive support, right skewness (most events resolve quickly), and heavy right tail (which captures rare but extreme values).
We model the mean structure ${\rm E}[{\ln}(D_n)]$ quadratically in ${\ln}(n)$ as:
\begin{align}
\label{eq:restorationMean}
    \mu(n) = {\ln}(\alpha_0) + \beta_1 {\ln}(n) + \beta_2 ({\ln}(n))^2
\end{align}
The quadratic term in \eqref{eq:restorationMean} allows the model to capture non-linear growth in restoration time and the saturation effects at very large event sizes. (A simple linear model was also tested, but the diagnostic plots showed systematic curvature in $\mu(n)$ estimates, which is resolved by the quadratic model.)
The intercept term $\alpha_0$ is the baseline restoration duration, $\beta_1$ is the primary power law exponent (slope of linear trend on log-log plot), and $\beta_2$ is the quadratic coefficient to control the curvature.
Since $D_n$ is lognormal, 
\begin{equation}
    {\rm Median}[D_n] = e^{\mu(n)} = \alpha_0\,
    n^{\beta_1} 
    n^{\beta_2{\ln}(n)}
\end{equation}
which means that the median has a power-law relationship with event size, with its exponent changing with event size depending on the quadratic term exponent $\beta_2$.

We note in Fig.~\ref{fig:restoreDurationVsEventSize} that the variability in restoration duration is not constant across event sizes. In particular, relative variability appears to decrease with increasing event size. 
Assuming a constant variance (homoskedastic errors) in such a situation distorts both the tail behavior and uncertainty estimates. Therefore, we use a heteroskedastic variance model \cite{carrollBook95} to allow variance to decay exponentially with event size:
\begin{align}
\label{eq:restorationSigma}
    \sigma(n) = \gamma_0 + \gamma_1 e^{-\delta n}
\end{align}
The exponential decay model \eqref{eq:restorationSigma} aligns with the change in variability shown in empirical data for all utilities in Fig.~\ref{fig:restoreDurationVsEventSize}. It captures the rapid initial decay in variability via the decay rate parameter $\delta$, allows additional variability $\gamma_1$ for small events, and ensures a minimum asymptotic variability $\gamma_0$. 
The coefficient of variation of $D_n$ 
can be calculated from the fact that it is a lognormal distribution as:
\begin{equation}
\label{eq:coefficientOfVariation}
    {\rm CV} = \sqrt{e^{\sigma^2}-1}
\end{equation}
As $n$ becomes large, \eqref{eq:coefficientOfVariation} becomes $\sqrt{e^{\gamma_0^2}-1}$. The marginal distribution of $D$ is obtained by integrating out $N$ as:
\begin{align}
    f_{D}(x) &= \sum_{n=2}^\infty f_{D_n}(x) \cdot \mathbb{P}(N=n) \\
     &= \sum_{n=2}^\infty \frac{\mathbb{P}(N=n)}{x\sigma(n)\sqrt{2\pi}}{\rm exp}(-\frac{({\ln}(x) - \mu(n))^2}{2\sigma^2(n)}) \label{eq:mixtureDistribution}
\end{align}
Since event size $N$ is discrete, the marginal distribution of $D$ is a mixture of lognormals \eqref{eq:mixtureDistribution}, naturally generating the heavy tails seen in Fig.~\ref{fig:restorationDuration}.

\subsection{Modeling the Time to First Restore}
\begin{figure}[t]
    \centering
    \includegraphics[width=1.0\linewidth]{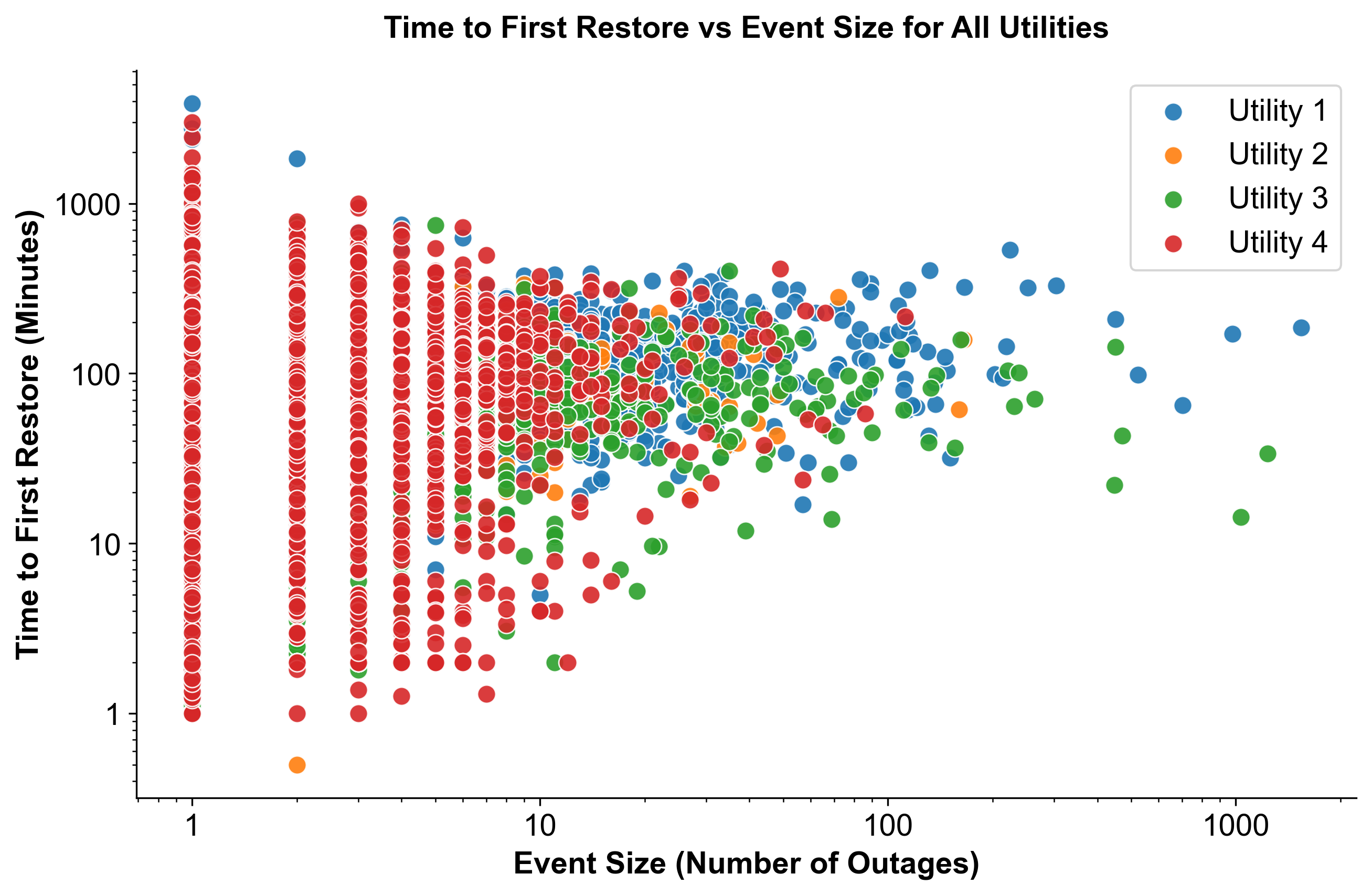}
    \caption{Variation of Time To First Restore with Event Size (log-log scale).}
    \label{fig:timeToFirstRestoreVsEventSize}
\end{figure}

The time to first restore $D_{r1}=r_1-o_1$ of an event is the difference between the time when a restore happens for the first time $r_1$ in that event and the start time $o_1$ of the first outage in the event.
If we exclude outages restored by the automatic operation of the distribution protection system, the time to first restore comprises the crew response time, travel time, and the time to repair the faulted component.
Unlike restoration durations, the empirical distributions of the time to first restore do not exhibit very heavy tails\footnote{Slope magnitudes of the distribution tails on log-log plots are 2.76, 1.78, 4.40, 3.52 for utilities 1 to 4, respectively.}, except for utility 2. 
Also, the time to first restore does not show a statistically significant dependence on event size; Fig.~\ref{fig:timeToFirstRestoreVsEventSize} visualizes the relationship. 
However, we observe two regimes with distinct variance trends: events with fewer than $\approx10$ outages have significantly higher variability (1 minute to 2.7 days), which decreases with event size, whereas events with $\approx10$ or more outages have relatively lower, more consistent variability (values centered around 100 minutes).
Small events inherently have higher variability because they may involve simple fixes such as breaker resets or fuse replacements, as well as slower cases involving hard-to-find problems, remote locations, and complex diagnosis/repair.
In contrast, the time to first restore for large events is relatively consistent because of defined procedures and protocols for responding to them.

Continuous distributions, including Lognormal, Gamma, and Weibull, are tested, but none provided a statistically significant best fit for the time to first restore across all events. Therefore, events with at least 10 outages are selected, and the Gamma distribution is found to provide the best fit. For events with fewer than 10 outages, the Lognormal and Gamma distributions give good fits for different utilities.

\section{Results and Discussion}

\subsection{Restore Times - Global Model}

\begin{figure}[t]
    \centering
    \includegraphics[width=1.0\linewidth]{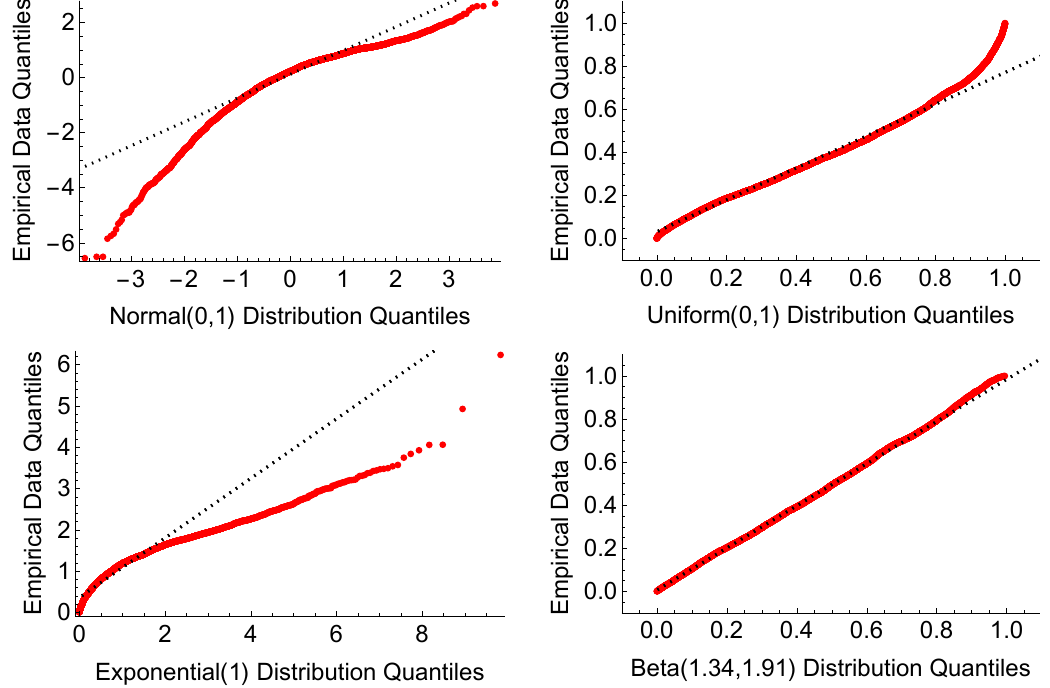}
    \caption{Quantile–quantile plots comparing empirical normalized restore times $\Delta r_i$ (Utility 1, events with $n \geq 30$) against candidate distributions.}
    \label{fig:globalFitComparison}
\end{figure}

\begin{figure}[t]
    \centering
    \includegraphics[width=1.0\linewidth]{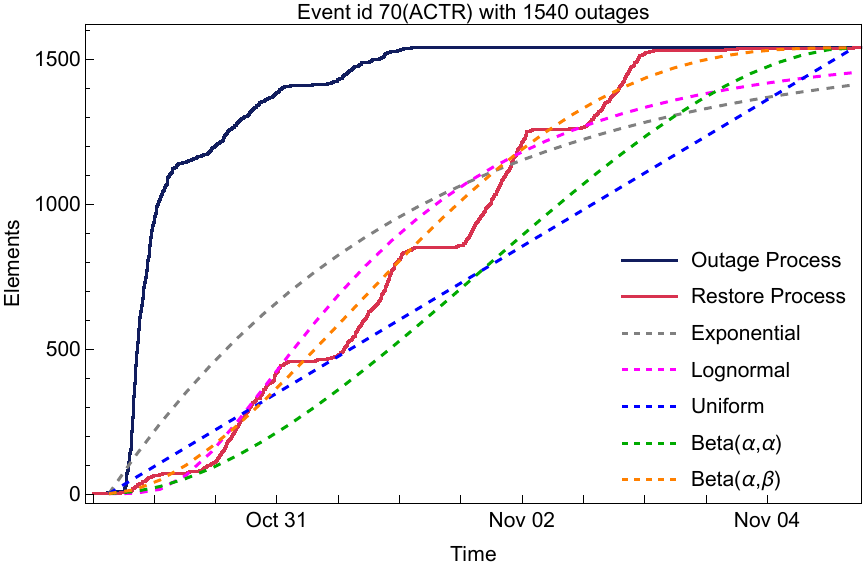}
    \caption{Example event restore process. Empirical outage and restore processes are compared with fitted restore time process models.}
    \label{fig:singleEventPlot}
\end{figure}

\looseness=-1
Fig.~\ref{fig:globalFitComparison} shows the quantile plots of restore times of events in Utility 1 data.
To ensure a sufficient number of data points, we select events with at least 30 outages for restore times modeling.
The results show that exponential and lognormal distributions are poor fits for the restore times, and the uniform and beta distributions are good candidates; the uniform provides a poor fit in the tails, while the beta provides a good fit overall. 
To substantiate this further, we look at the fits of these candidate distributions to the largest event in the data in Fig.~\ref{fig:singleEventPlot}. We can see that, according to the lognormal model, the restore times slow down significantly towards the end of the event, thereby failing to accurately model the restore process. The exponential distribution performs poorly in modeling both the start and the end of the restore process. The uniform distribution gives a good fit in an average sense, whereas the restore times per the beta distribution nicely align with the empirical restore times.
This is because the beta distribution fits the upper and lower tails of the data better (due to its inverted U shape, it gives low probability in the tails).
Based on these qualitative results, we rule out the lognormal and exponential distributions and proceed to test the beta and uniform distributions quantitatively.
We note in passing that the beta distribution captures the diurnal patterns in which repairs sharply decrease at night in Fig.~\ref{fig:singleEventPlot} only in an averaged sense.

\begin{figure*}[ht]
    \centering
    \includegraphics[width=\textwidth]{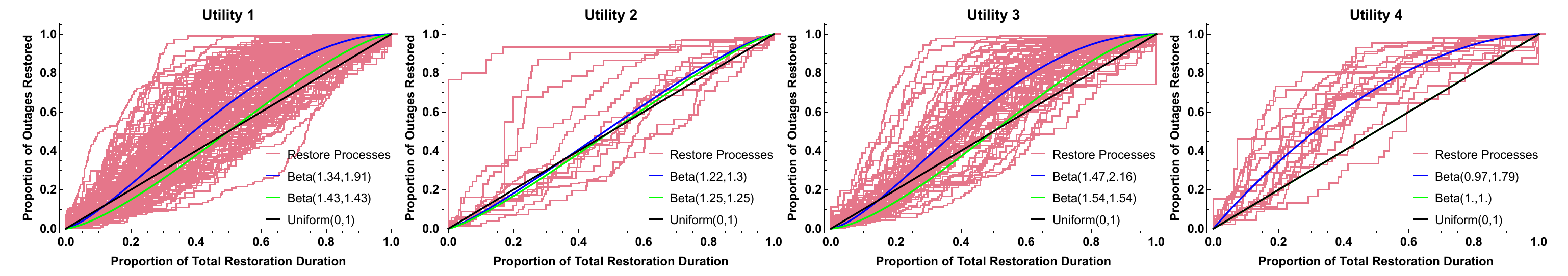}
    \caption{Global fitted models of restore times overlaid on the normalized empirical restore processes of events with size $n \geq 30$.}
    \label{fig:allUtilities}
\vspace{-1em}
\end{figure*}

Fig.~\ref{fig:allUtilities} shows the global models with uniform and beta distributions overlaid on the restore curves for all events (with at least 30 outages) across the four utilities. 
Table~\ref{table:globalResults} shows the best fit parameters along with goodness-of-fit metrics. A more negative AICc value indicates a better model, based on both the likelihood and the model complexity.
Based on these results, we conclude that the uniform distribution and the single-parameter beta distribution give comparable fits across all four cases, whereas the two-parameter beta distribution provides an overall superior fit, especially in tail regions.

\begin{table}[hbpt]
	\caption{Results of Global Models of Restore Times} 
	\label{table:globalResults}
	\centering
\begin{tabular}{ l @{\hspace{-1pt}} c c c c}
                        &Utility-1  &Utility-2  &Utility-3  &Utility-4  \\
\midrule
\textbf{Standard Uniform} &      &     &      &            \\
KS Distance             & 0.16      & 0.07      & 0.19      & 0.28          \\    [3pt]

\textbf{Beta($\alpha,\alpha$)} Parameters $\quad$  & (1.4,1.4)  & (1.3,1.3) & (1.5,1.5) & (1.0,1.0)          \\
KS Distance             & 0.15      & 0.05      & 0.18      & 0.28          \\    
AICc                    & -834      & -19       & -741      & 2          \\[3pt]

\textbf{Beta($\alpha,\beta$)} Parameters   & (1.3,1.9)  & (1.2,1.3) & (1.5,2.2) & (1.0,1.8)          \\
KS Distance             & 0.01      & 0.05       & 0.02       & 0.07          \\    
AICc                    & -2198     & -20        & -1863      & -236          \\
\end{tabular}
\end{table}

\looseness=-1
We note that in the two-parameter beta model, $\beta>\alpha$ for all utilities, indicating that the restoration process is ``front-loaded" with the probability density more concentrated at the beginning.
For example, the mean is $\alpha/(\alpha+\beta)\approx0.40$ for utility~1, indicating that a significant portion of the components are restored within the first 40\% of the total restoration duration window, while the remaining customers take disproportionately longer to fix. This implies that the repair rate decreases over time, likely because the easy fixes and automatic restorations are done first, leaving the complex ones for the tail end.

\subsection{Restore Times - Individual Model}
The goodness-of-fit of the uniform and beta distributions is tested for the restore process of each event with 30 or more outages. Since these are nested distributions with increasing complexity, the likelihood ratio test is used instead of AICc to quantify the goodness-of-fit and its trade-off with complexity. 
The comparison of Uniform(0,1) and Beta($\alpha,\alpha$) in Table~\ref{table:gofRestoreTimes} shows that the Beta distribution gives better AICc scores for all utilities.
However, this better fit is due to an additional model parameter, as reflected in the Likelihood ratio test results (except for Utility 3). The LRT shows that, at the 0.05 significance level, in most cases the more complex Beta($\alpha,\alpha$) model is unnecessary and the Uniform distribution is sufficient to model restore times for individual events.

\begin{table}[hbpt]
	\caption{Results of Individual Models of Restore Times, showing the percentage of  events for which Distribution-A scored better than Distribution-B for different GOF criteria} 
	\label{table:gofRestoreTimes}
	\centering
\begin{tabular}{ l @{\hspace{-1pt}} c c c c c}
                        &Utility-1  &Utility-2  &Utility-3  &Utility-4  \\
\midrule
\multicolumn{5}{l}{\textbf{Dist. A} = Beta$(\alpha, \alpha)$, \textbf{Dist. B} = Uniform$(0, 1)$}\\
KS Distance             & 67\%      & 83\%       & 77\%       & 69\%         \\    
LRT                     & 49\%      & 42\%        & 61\%      & 31\%          \\ [3pt]
\multicolumn{5}{l}{\textbf{Dist. A} = Beta$(\alpha, \beta)$, \textbf{Dist. B} = Beta$(\alpha, \alpha)$}\\
KS Distance             & 94\%      & 92\%       & 94\%       & 100\%          \\    
LRT                     & 55\%      & 50\%       & 63\%         & 75\%          \\
\end{tabular}
\end{table}

The LRT results in the lower half of Table~\ref{table:gofRestoreTimes} suggest that the additional complexity of the Beta($\alpha,\beta$) distribution is statistically significant as it fits the data better than the Beta($\alpha,\alpha$) distribution for more than 50\% of the events.

\begin{figure*}[t]
    \centering
    \includegraphics[width=\textwidth]{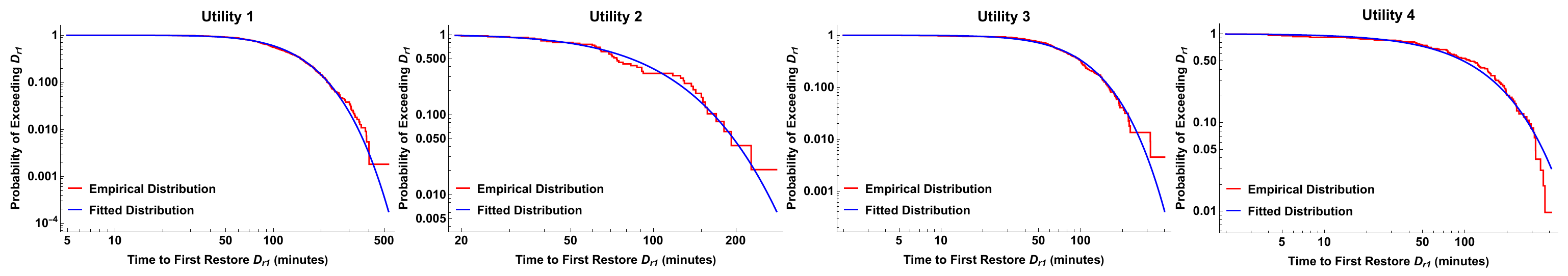}
    \caption{Empirical CCDF (red) along with the fitted Gamma Distribution (blue) of Time To First Restore of events with at least 10 outages.}
    \label{fig:timeToFirstRestore}
\vspace{-1em}
\end{figure*}

\subsection{Event Restoration Duration}
We use maximum likelihood to estimate parameters of \eqref{eq:restorationDurationModel}, \eqref{eq:restorationMean}, and \eqref{eq:restorationSigma}. Given the data $\{(x_1,n_1),...(x_j,n_j)\}$ and the parameter vector $\theta=(\alpha_0,\beta_1,\beta_2,\gamma_0,\gamma_1,\delta)$, we substitute the lognormal density of \eqref{eq:restorationDurationModel} in the following log-likelihood and maximize it over $\theta$ to estimate the parameters subjected to $\alpha_0 >0, \gamma_0 \geq 0, \gamma_1 \geq 0$:
\begin{equation}
    l(\theta) = \sum_{i=1}^j {\ln}f_{D_n}(x_i,n_i;\theta)
\end{equation}
The estimated parameter values are tabulated in Table~\ref{tab:restoration_duration_results}.
Utility 2 has the minimum baseline restoration duration $\alpha_0$ for events with 2 outages.
All utilities have $\beta_1>1$, which means restoration duration scales superlinearly with event size. This suggests coordination and scalability challenges during large events.
The small, negative quadratic curvature coefficient $\beta_2$ indicates that scaling becomes slightly less steep for large events, and the marginal effect of additional outages on the restoration duration decreases.
A lower $\gamma_0$ value is better as it governs the irreducible uncertainty for very large events. For example, for utility 4, the coefficient of variation \eqref{eq:coefficientOfVariation} is 98\%, indicating that very large events exhibit high variability, which makes it more difficult to accurately estimate the restoration duration.
Additional variability for small events, in addition to the minimum variability $\gamma_0$, is captured by $\gamma_1$. 
Utility 2 has the smallest ratio of $\gamma_1/\gamma_0 = 1.6$, which means the small and large events have similar variability in restoration duration.
The variability decay rate $\delta$ is the largest for utility 4, which means that the small event variability is gone by $n=10$, i.e., $2.58e^{-0.65\times10}\approx0$.

\begin{table}[hbpt]
	\caption{Parameter estimates of the Event Restoration Duration model} 
	\label{tab:restoration_duration_results}
	\centering
\begin{tabular}{ l @{\hspace{-1pt}} c c c c c}
       \qquad{} &Utility-1  &Utility-2  &Utility-3  &Utility-4  \\
\midrule
$\alpha_0$ (minutes)        & 26.14     & 14.03     & 17.04     & 20.16         \\    
$\beta_1$       & 1.45      & 1.35      & 1.60      & 1.92          \\
$\beta_2$       & -0.10     & -0.05     & -0.13     & -0.20          \\
$\gamma_0$      & 0.42      & 0.74      & 0.48      & 0.85          \\  
$\gamma_1$      & 1.89      & 1.18      & 1.29      & 2.58          \\
$\delta$        & 0.42      & 0.25      & 0.26      & 0.65          \\
\end{tabular}
\vspace{-1em}
\end{table}

\subsection{Time to First Restore}
Fig.~\ref{fig:timeToFirstRestore} shows the empirical CCDF of the time to first restore for events with at least 10 outages, along with the fitted Gamma distributions. 
Restricting to events with at least 10 outages reduces the high-variance regime visible in Fig.~\ref{fig:timeToFirstRestoreVsEventSize} and yields a more stable distribution across utilities. 
The fitted Gamma parameters are shown in Table~\ref{tab:timeToFirstRestoreParams}. 
Utilities 1–3 have similar scale parameters ($\theta \approx 32-38$ minutes) and moderate shape parameters ($k \approx 2.4-3.4$), which correspond to typical times to first restore, $k \theta$, on the order of 1–2 hours. Utility 4 differs primarily in its variability: its fitted shape, $k = 1.2$, implies substantially higher dispersion, and its larger scale ($\theta \approx 104$ minutes) shifts the distribution to longer times. 
The coefficient of variation is $1/\sqrt{k}$, which is lowest for Utility 1 (more consistent time to first restore) and highest for Utility 4 (least predictable time to first restore).

\begin{table}[hbpt]
	\caption{Estimated parameters of Time to First Restore (minutes) fit} 
	\label{tab:timeToFirstRestoreParams}
	\centering
\begin{tabular}{ l @{\hspace{-1pt}} c c c c c}
                        &Utility-1  &Utility-2  &Utility-3  &Utility-4  \\
\midrule
Gamma Shape $k$\quad{} & 3.35      & 2.95      & 2.43      & 1.24          \\
Gamma Scale $\theta$        & 38.25     & 31.58     & 35.97     & 103.63          
\end{tabular}
\end{table}

\section{Conclusion}

This paper develops data-driven probabilistic models for three components of the distribution system restoration process: normalized restore-time progression, total restoration duration, and time to first restore. 
We identify model classes that fit reliably across utilities using data from four US utilities.

Beyond identifying best-fit distribution classes, the models provide an operationally interpretable representation of restoration. 
In practice, once the outage process of an event ends, the utility soon knows the number of outages (the event size). The conditional lognormal duration model~\eqref{eq:restorationDurationModel} of restoration duration enables probabilistic forecasting of total restoration duration from event size, supporting early resource allocation, escalation decisions, and communication of expected downtime. 
The restore-time process model for $\Delta r_i$ provides a typical ``shape" of restoration progression within the event window, which is not captured by common practice that relies on means/medians or assumes exponential repair behavior. 
By presenting the uniform distribution as a parsimonious alternative and the Beta($\alpha,\beta$) distribution as the best-fit model for restore times of events with $n\geq30$, the results support selecting a model for low- and high-fidelity simulations that remains statistically defensible.
These results on typical restore times in distribution systems differ from those in transmission systems \cite{DobsonPS24}, where a lognormal model is used rather than a uniform or beta distribution.
Time to first restore of moderate and large events shows a weaker dependence on event size than total restoration duration, but it exhibits a clear variance-regime change: small events have much higher variability than large events.
Utility-to-utility differences in the Gamma parameters indicate that the time to first restore is more utility-specific than the normalized restore-time process shape, which is comparatively stable across utilities.
Together, these models can support uncertainty-aware Monte Carlo simulation, benchmarking, and restoration-duration forecasting using utility outage data.

\printbibliography

\end{document}